\documentstyle[12pt]{article}
\newcommand{\be}{\begin{equation}}
\newcommand{\ee}{\end{equation}} 
\newcommand{\eei}{\end{equation}\indent\indent}
\newcommand{\bc}{\begin{center}}
\newcommand{\ec}{\end{center}}
\newcommand{\ber}{\begin{eqnarray}}
\newcommand{\ear}{\end{eqnarray}}
\newcommand{\ba}{\begin{array}}
\newcommand{\ea}{\end{array}}

\def\case#1/#2{\textstyle\frac{#1}{#2} }

\catcode`\@=11
\long\def\@makefntext#1{
\protect\noindent \hbox to 3.2pt {\hskip-.9pt
$^{{\ninerm\@thefnmark}}$\hfil}#1\hfill}		

 \def\@makefnmark{\hbox to 0pt{$^{\@thefnmark}$\hss}}  

\def\ps@myheadings{\let\@mkboth\@gobbletwo
\def\@oddhead{\hbox{}
\rightmark\hfil\ninerm\thepage}
\def\@oddfoot{}\def\@evenhead{\ninerm\thepage\hfil
\leftmark\hbox{}}\def\@evenfoot{}
\def\sectionmark##1{}\def\subsectionmark##1{}}

\newcounter{sectionc}\newcounter{subsectionc}\newcounter{subsubsectionc}
\renewcommand{\section}[1] {\vspace{0.6cm}\addtocounter{sectionc}{1}
\setcounter{subsectionc}{0}\setcounter{subsubsectionc}{0}\noindent
 	{\bf\thesectionc. #1}\par\vspace{0.4cm}}
\renewcommand{\subsection}[1] {\vspace{0.6cm}\addtocounter{subsectionc}{1}
 	\setcounter{subsubsectionc}{0}\noindent
 	{\it\thesectionc.\thesubsectionc. #1}\par\vspace{0.4cm}}
\renewcommand{\subsubsection}[1] {\vspace{0.6cm}\addtocounter{subsubsectionc}{1}
  	\noindent {\rm\thesectionc.\thesubsectionc.\thesubsubsectionc.
  	#1}\par\vspace{0.4cm}}

\newcounter{appendixc}
\newcounter{subappendixc}[appendixc]
\newcounter{subsubappendixc}[subappendixc]

\renewcommand{\appendix}[1] {\vspace{0.6cm}
        \refstepcounter{appendixc}
        \setcounter{figure}{0}
        \setcounter{table}{0}
        \setcounter{equation}{0}
        \renewcommand{\thefigure}{\Alph{appendixc}.\arabic{figure}}
        \renewcommand{\thetable}{\Alph{appendixc}.\arabic{table}}
        \renewcommand{\theappendixc}{\Alph{appendixc}}
        \renewcommand{\theequation}{\Alph{appendixc}.\arabic{equation}}
        \noindent{\bf Appendix \theappendixc #1}\par\vspace{0.4cm}}

\def\abstracts#1{{
 	\centering{\begin{minipage}{30pc}\tenrm\baselineskip=12pt\noindent
 	\centerline{\tenrm ABSTRACT}\vspace{0.3cm}
 	\parindent=0pt #1
 	\end{minipage}}\par}}

\renewenvironment{thebibliography}[1]
 	{\begin{list}{\arabic{enumi}.}
 	{\usecounter{enumi}\setlength{\parsep}{0pt}
\setlength{\leftmargin 1.25cm}{\rightmargin 0pt}
 	 \setlength{\itemsep}{0pt} \settowidth
 	{\labelwidth}{#1.}\sloppy}}{\end{list}}

\newcounter{itemlistc}
\newcounter{romanlistc}
\newcounter{alphlistc}
\newcounter{arabiclistc}

\newcommand{\fcaption}[1]{
        \refstepcounter{figure}
        \setbox\@tempboxa = \hbox{\tenrm Fig.~\thefigure. #1}
        \ifdim \wd\@tempboxa > 6in
           {\begin{center}
        \parbox{6in}{\tenrm\baselineskip=12pt Fig.~\thefigure. #1}
            \end{center}}
        \else
             {\begin{center}
             {\tenrm Fig.~\thefigure. #1}
              \end{center}}
        \fi}

\newcommand{\tcaption}[1]{
        \refstepcounter{table}
        \setbox\@tempboxa = \hbox{\tenrm Table~\thetable. #1}
        \ifdim \wd\@tempboxa > 6in
           {\begin{center}
        \parbox{6in}{\tenrm\baselineskip=12pt Table~\thetable. #1}
            \end{center}}
        \else
             {\begin{center}
             {\tenrm Table~\thetable. #1}
              \end{center}}
        \fi}

\def\@citex[#1]#2{\if@filesw\immediate\write\@auxout
 	{\string\citation{#2}}\fi
\def\@citea{}\@cite{\@for\@citeb:=#2\do
 	{\@citea\def\@citea{,}\@ifundefined
 	{b@\@citeb}{{\bf ?}\@warning
 	{Citation `\@citeb' on page \thepage \space undefined}}
 	{\csname b@\@citeb\endcsname}}}{#1}}

\newif\if@cghi
\def\cite{\@cghitrue\@ifnextchar [{\@tempswatrue
 	\@citex}{\@tempswafalse\@citex[]}}
\def\citelow{\@cghifalse\@ifnextchar [{\@tempswatrue
 	\@citex}{\@tempswafalse\@citex[]}}
\def\@cite#1#2{{$\null^{#1}$\if@tempswa\typeout
 	{IJCGA warning: optional citation argument
 	ignored: `#2'} \fi}}

\def\fnt#1#2{\footnotetext{\kern-.3em
 	{$^{\mbox{\sevenrm #1}}$}{#2}}}

 1
 1
 1

\font\tenbf=cmbx10
\font\tenrm=cmr10
\font\tenit=cmti10

\font\ninerm=cmr9

\begin{document}

\centerline{\tenbf ON RELATIVISTIC PERTURBATIONS OF SECOND AND HIGHER ORDER }
\vspace{0.8cm}
\centerline{\tenrm MARCO BRUNI}
\baselineskip=13pt
\centerline{\tenit 
ICTP, International Center for Theoretical Physics, P. O. Box 586,
34014 Trieste, Italy}
\centerline{\tenit SISSA, International School for Advanced Studies, 
via Beirut 2-4, 34013 Trieste, Italy}

\vspace{0.3cm}
\centerline{\tenrm SABINO MATARRESE}
\baselineskip=13pt
\centerline{\tenit  Dipartimento di Fisica ``Galileo Galilei",}
\centerline{\tenit
Universit\`a di Padova, via Marzolo 8, 35131 Padova, Italy}
\vspace{0.3cm}
\centerline{\tenrm SILVIA MOLLERACH and SEBASTIANO SONEGO}
\centerline{\tenit SISSA, International School for Advanced Studies, 
via Beirut 2-4, 34013 Trieste, Italy}

\vspace{0.9cm}
\abstracts{
We present the results of a study of the gauge dependence
of spacetime perturbations. In particular, we consider gauge
invariance in general, we give a generating
formula for gauge transformations to an arbitrary order $n$, and
explicit transformation rules at second order.
}
\bigskip\bigskip\bigskip
\bigskip\bigskip\bigskip
\centerline{\it Proceedings of the }
\centerline{\it 12th Italian Conference on General
Relativity and Gravitational Physics}
\bigskip\bigskip\bigskip
\bigskip\bigskip\bigskip
\bigskip\bigskip\bigskip
\centerline{IC/96/224}
\centerline{SISSA--162/96/A}
\newpage
\rm\baselineskip=14pt
\section{Introduction}

Second order treatments have been recently proposed, both in
cosmology\cite{bi:MPS,bi:russ}  and
compact object theory,\cite{bi:gleiseretal} 
 as a way of obtaining more accurate results to
be compared with present and future observations. Also, second order
perturbations provide a reliable measure of the accuracy of the
linearized theory. However, as it is well known, relativistic
perturbations are,  in general, gauge dependent.\cite{bi:sachs,bi:SW}
 Here we illustrate
some results we have  recently derived,\cite{bi:BMMS}
 concerning this issue and the one 
of gauge invariance.\cite{bi:SW,bi:bardeen,bi:EB}  We 
shall omit proofs, for reasons of space.

\section{Knight diffeomorphisms}

Let ${\cal M}$ be a $m$-dimensional 
differentiable manifold, and let $\xi$ be a vector
field on ${\cal M}$, generating a flow\footnote{In order not to burden 
the discussion unnecessarily, we suppose that $\phi$ is a one-parameter 
group of diffeomorphisms, defining global transformations 
of $\cal M$.} $~\phi:{\rm I\!R}\times{\cal M}\to{\cal M}$, where $\phi(0,p)=p$,
$\forall\,p\in{\cal M}$. For any given $\lambda\in{\rm I\!R}$, we
shall write, as  usual,
$\phi_\lambda(p):=\phi(\lambda,p)$, $\forall\, p\in{\cal M}$.  If $T$
is a tensor field on ${\cal M}$, the pull--back $\phi_{\lambda}^*$ defines a
new field $\phi^*_{\lambda} T$ on ${\cal M}$, 
which is thus a function of $\lambda$. Then it is known\cite{bi:schouten} 
that $\phi^*_{\lambda} T$ 
admits the following expansion around $\lambda=0$:
\begin{equation}
 \phi^*_{\lambda} T=\sum^{+\infty}_{k=0}
\,\frac{\lambda^k}{k!}\,\pounds^k_\xi T={\rm e}^{\lambda\pounds_\xi}T\;,
\label{lemma1}
\end{equation}
where $\pounds_\xi$ denotes the Lie derivative along $\xi$.  It is worth   
pointing out that the proof of (\ref{lemma1}) uses the  group property 
$\phi_{\lambda+\sigma}=\phi_\lambda \circ \phi_\sigma$.

Let us now suppose that there are {\em two\/} vector fields
$\xi_{(1)}$ and $\xi_{(2)}$ on ${\cal M}$,  generating
the flows $\phi^{(1)}$ and $\phi^{(2)}$.  We can combine
$\phi^{(1)}$ and $\phi^{(2)}$ to define a new one-parameter family of
diffeomorphisms $\Psi:{\rm I\!R}\times{\cal M}\to{\cal M}$, whose
action is given by $\Psi_\lambda :=
\phi^{(2)}_{\lambda^2/2}\circ\phi^{(1)}_\lambda$. Thus, $\Psi_\lambda$
displaces a point of ${\cal M}$ a parameter interval $\lambda$ along
the integral curve of $\xi_{(1)}$, and then an interval $\lambda^2/2$
along the integral curve of $\xi_{(2)}$.
With a chess-inspired terminology,
we shall call it a {\em knight diffeomorphism},  or simply a {\em knight}.
This concept can be immediately generalized to the case in which $n$
vector fields $\xi_{(1)},\ldots,\xi_{(n)}$ are defined on ${\cal M}$,
corresponding to the flows $\phi^{(1)},\ldots, \phi^{(n)}$.  Then we
define a one-parameter family $\Psi:{\rm I\!R}\times{\cal M}\to{\cal
M}$ of knights of rank $n$ by
\begin{equation}\label{eq:knightn}
\Psi_\lambda:=\phi^{(n)}_{\lambda^n/n!}
\circ\cdots\circ\phi^{(2)}_{\lambda^2/2}
\circ\phi^{(1)}_\lambda\;,
\end{equation}
and the vector fields $\xi_{(1)},\ldots,\xi_{(n)}$ will be called the
{\em generators\/} of $\Psi$.
Of course, $\Psi_\sigma\circ\Psi_\lambda\neq\Psi_{\sigma+\lambda}$;
consequently,  (\ref{lemma1}) cannot be applied if we want to expand in
$\lambda$ the pull-back $\Psi_{\lambda}^* T$ of a tensor field $T$
defined on ${\cal M}$.  However, the result is  easily
extended, because the pull-back 
$\Psi_{\lambda}^* T$ of a tensor field $T$ by a
one-parameter family of knights
$\Psi$ with generators $\xi_{(1)},\ldots,\xi_{(k)},
\ldots$ can be expanded
around $\lambda=0$ as follows: 
\begin{equation}
\Psi_{\lambda}^* T=\sum_{l_1=0}^{+\infty}
 \sum_{l_2=0}^{+\infty}\cdots
\sum_{l_k=0}^{+\infty}\cdots \,
 {\lambda^{l_1+2l_2+\cdots+kl_k+\cdots}\over
2^{l_2}\cdots (k!)^{l_k}\cdots l_1!l_2!\cdots
l_k!\cdots}\,\pounds^{l_1}_{\xi_{(1)}}\pounds^{l_2}_{\xi_{(2)}}
\cdots\pounds^{l_k}_{\xi_{(k)}}\cdots
T\;.
\label{lemma2}
\end{equation}
The proof of this simply requires the repeated application of
 (\ref{lemma1}) .
The explicit form of (\ref{lemma2}) up to the second order in
$\lambda$ is
\begin{equation}
\Psi_{\lambda}^* T=T+\lambda \pounds_{\xi_{(1)}} T
+\frac{\lambda^2}{2}\left(\pounds^2_{\xi_{(1)}}
+\pounds_{\xi_{(2)}}\right)T+\cdots\;.
\label{lemma2explic}
\end{equation}
Equations (\ref{lemma2}) and (\ref{lemma2explic}) apply to a
one-parameter family of knights  of arbitrarily high
rank, and can be specialized to the particular case of rank $n$ simply
by setting $\xi_{(k)}\equiv 0$, $\forall\, k>n$.  
Applying   (\ref{lemma2})  to one of the coordinate functions on $\cal M$,
$x^\mu$, we have, since
$\Psi^*_{\lambda}x^\mu(p)=x^\mu(\Psi_\lambda(p))$, the extention  to
second order in $\lambda$ of the  action of
an ``infinitesimal point transformation'':
\begin{equation} 
\tilde{x}^\mu:=x^\mu(\Psi_\lambda(p))=x^\mu(p)
+\lambda\,\xi_{(1)}^\mu+{\lambda^2\over
2}\,\left({\xi_{(1)}^\mu}_{,\nu}\xi_{(1)}^\nu
+\xi_{(2)}^\mu\right)+\cdots\;.
\label{lemma2coord}
\end{equation}

Knights are rather special, and (\ref{lemma2}) may seem of limited
applicability.  This is, however, not the case, as shown by the
following

\noindent {\bf Theorem:} {\it Let $\Psi:{\rm I\!R}\times{\cal M}\to {\cal
M}$ be a one-parameter family of diffeomorphisms.  Then $\exists$
$\phi^{(1)},\ldots,\phi^{(k)},\ldots,$ one-parameter groups of
diffeomorphisms of $\cal M$, such that}
\begin{equation}
\Psi_\lambda=\cdots\circ\phi^{(k)}_{\lambda^k/k!}\circ\cdots
\circ\phi^{(2)}_{\lambda^2/2}\circ\phi^{(1)}_\lambda\;.
\label{theorem}
\end{equation}
The meaning of this Theorem is  that any one-parameter family of
diffeomorphisms can always be regarded as a one-parameter family of
knights --- of infinite rank, in general --- and can be approximated
by a family of knights of suitable rank.
\footnote{ We have supposed so
far that maps and fields are analytic, but it is possible to give
versions of (\ref{lemma1}), (\ref{lemma2}), and (\ref{theorem}), that
hold only for $C^n$ objects. The main change\cite{bi:SB} then is the
substitution of Taylor series like the one in (\ref{lemma1}) by a
finite sum of $n-1$ terms plus a remainder.}

\section{Gauge transformations}

Consider now  a family  of spacetime models
$\{({\cal M},g_\lambda,\tau_\lambda)\}$, where the metric $g_\lambda$ and
the matter fields (here collectively referred to as $\tau_\lambda$)
satisfy the field equation \hbox{${\cal E}[g_\lambda,\tau_\lambda]=0$}, 
and $\lambda\in{\rm I\!R}$.  We  assume that $g_\lambda$ and
$\tau_\lambda$ depend smoothly on the dimensionless parameter 
$\lambda$, so that $\lambda$ itself is a measure of the amount by which a 
specific $({\cal M},g_\lambda,\tau_\lambda)$ differs from the 
background solution $({\cal M},g_0,\tau_0)$, which is supposed to
be known. 
This situation is most naturally described by introducing an
$(m+1)$-dimensional manifold $\cal N$, foliated by submanifolds
diffeomorphic to $\cal M$, so that ${\cal N}={\cal M}\times{\rm
I\!R}$.  We shall label each copy of $\cal M$ by the corresponding
value of the parameter $\lambda$.  
Now, if a tensor field $T_\lambda$ is given on each ${\cal
M}_\lambda$,  a tensor field $T$ is automatically defined
on ${\cal N}$.

  In order to define the perturbation in $T$, we
must find a way to compare $T_\lambda$ with $T_0$. 
This requires a prescription for identifying points of ${\cal
M}_\lambda$ with those of ${\cal M}_0$,  which is given by 
a diffeomorphism $\varphi_\lambda:{\cal N}\to{\cal N}$
such that $\left.\varphi_\lambda\right|_{\scriptscriptstyle {\cal
M}_0}:{\cal M}_0\to{\cal M}_\lambda$.  Clearly, $\varphi_\lambda$ can
be regarded as the member of a flow $\varphi$ on $\cal N$,
corresponding to the value $\lambda$ of the group parameter.
Therefore, we could equally well give the vector field $X$ that
generates $\varphi$,  and we shall refer both to the point 
identification map $\varphi$ and to  $X$ as a {\em gauge 
choice\/}. The perturbation can now be defined simply as
\begin{equation}
\Delta T_\lambda:=\left.\varphi^*_\lambda T\right|_{\scriptscriptstyle
{\cal M}_0}-T_0\;.
\label{DeltaT}
\end{equation}
The first term on the right hand side of  (\ref{DeltaT}) can be
Taylor-expanded to get\footnote{For the sake of simplicity, we 
 denote the restriction to ${\cal M}_0$ of a
tensor field defined over $\cal N$ simply by the suffix 0.}
\begin{equation}
\Delta T_\lambda=\sum_{k=1}^{+\infty}{\lambda^k\over k!}\,\delta^kT\;,
~~~~
\delta^kT:=\left[{{\rm d}^k\varphi^*_\lambda T\over {\rm
d}\lambda^k}\right]_{\lambda=0,{\cal M}_0}=\pounds^k_XT|_0\;.
\label{delta1T}
\end{equation}
Equation (\ref{delta1T}) defines then the $k$-th order perturbation of
$T$.  Notice that $\Delta T_\lambda$ and $\delta^k T$ are defined on
${\cal M}_0$; this formalizes the statement one commonly finds in the
literature, that ``perturbations are fields living in the
background.''  In the particular case when $T$ is the metric or the 
matter fields, $\delta^k g$ and $\delta^k \tau$ obey {\em linear\/} 
equations, obtained by differentiating ${\cal E}\left[ 
g_\lambda,\tau_\lambda\right]=0$ with respect to $\lambda$.  This gives 
an iterating procedure to calculate $g_\lambda$ and $\tau_\lambda$ when 
the field equation is too difficult to solve exactly.

Let us now suppose that {\it two} gauges $X$ and $Y$ are defined,
associated with  $\varphi$ and
$\psi$ on $\cal N$, that connect any two leaves of the foliation.  
Thus $X$ and
$Y$ are everywhere transverse to the ${\cal M}_\lambda$, and points
lying on the same integral curve of either of the two are to be regarded 
{\em as the same
point\/} within the respective gauge, i.e.,  $\varphi$ and $\psi$ are both 
point identification maps.
Both can  be used to pull back a generic tensor field
$T$, and to construct therefore two other tensor fields
$\varphi^*_{\lambda} T$ and $ \psi^*_{\lambda} T$, for any given
value of $\lambda$.  In particular, on ${\cal M}_0$ we now have
three tensor fields, i.e., $T_0$, and
\begin{equation}
\label{eq:txy}
T^X_\lambda := \left.\varphi^*_{\lambda} T\right|_0\, , ~~~
T^Y_\lambda := \left.\psi^*_{\lambda} T\right|_0\, . 
\end{equation}
Since  $X$ and $Y$ represent gauge choices for mapping a 
perturbed manifold ${\cal M}_\lambda$ onto the unperturbed one ${\cal
M}_0$, $T^X_\lambda$ and $T^Y_\lambda$ are the representations, in
${\cal M}_0$, of the perturbed tensor according to the two gauges.  We 
can write, using (\ref{DeltaT}), (\ref{delta1T}) and (\ref{eq:txy}),
\begin{equation} 
T^X_\lambda=\sum_{k=0}^{+\infty}\frac{\lambda^k}{k!}\,
\delta^k T^X =
\sum_{k=0}^{+\infty}{\lambda^k\over k!}\,
\left.\pounds^k_X T\right|_0\,
\;, ~~~~
T^Y_\lambda=\sum_{k=0}^{+\infty}\frac{\lambda^k}{k!}\,
\delta^k T^Y =
\sum_{k=0}^{+\infty}{\lambda^k\over k!}\,
\left.\pounds^k_Y T\right|_0\,
\;.
\label{4.4}
\end{equation}


If $T^X_\lambda=T^Y_\lambda$, for any pair of gauges $X$ and
$Y$, we  say that $T$ is {\em totally gauge-invariant\/}.  This
is a  strong condition, because then  (\ref{4.4}) 
 imply that $\delta^kT^X=\delta^kT^Y$, for all  $X$ and $Y$
and  $\forall k$.  But in practice, one is  interested in 
perturbations to order $n$;  
it is thus convenient to weaken the definition, saying that $T$
is {\em gauge-invariant to order\/} $n$ iff
$\delta^kT^X=\delta^kT^Y$ for any $X$ and
$Y$, and $\forall k\leq n$.  One can then prove the following

\noindent {\bf Proposition 1:} {\it A tensor field $T$ is 
gauge-invariant to order $n\geq 1$ iff $\pounds_\xi\delta^kT=0$, for
any vector field $\xi$ on $\cal M$ and $\forall k<n$.}

As a consequence, $T$ is gauge-invariant to order $n$ iff $T_0$ and all its
perturbations of order lower than $n$ are, in any gauge, 
 a combination of Kronecker deltas
with constant coefficients.\cite{bi:sachs,bi:SW} 
Further, it then
follows that $T$ is totally gauge-invariant iff  it
is  a combination of Kronecker deltas with coefficients depending only 
on $\lambda$.

If a tensor $T$ is not gauge-invariant, it is important to know how
its representation on ${\cal M}_0$ changes under a gauge
transformation.  To this purpose, it is useful to define, for each
value of $\lambda\,\in{\rm I\!R}$, the diffeomorphism
$\Phi_\lambda\,:\,{\cal M}_0\to {\cal M}_0$ given by 
$
\Phi_\lambda :=\varphi_{-\lambda}\circ\psi_{\lambda}\;.
$
We must stress that $\Phi:{\rm I\!R}\times{\cal
M}_0\to{\cal M}_0$ so defined, {\em is not\/} a flow 
on ${\cal M}_0$.  In fact,
$\Phi_{-\lambda}\not=\Phi^{-1}_\lambda$, and
$\Phi_{\lambda+\sigma} \not= \Phi_\sigma\circ\Phi_\lambda$, essentially
because  $X$ and $Y$, in general, do not 
commute.  However, the Theorem  above 
guarantees that, to order $n$ in $\lambda$, the one-parameter family of
diffeomorphisms $\Phi$ can always be approximated by a one-parameter
family of knights of rank $n$.
It is very easy to see that the tensor fields $T^X_\lambda$ and
$T^Y_\lambda$ defined by the gauges $\varphi$ and $\psi$ are connected
by the linear map $\Phi_\lambda^*$:
\begin{equation}
T^Y_\lambda=\left.\psi^*_{\lambda} T\right|_0=
\left.(\psi^*_{\lambda}\varphi^*_{-\lambda}\varphi^*_{\lambda}
T)\right|_0 =\left.\Phi_\lambda^*(\varphi^*_{\lambda}T)\right|_0
=\Phi_\lambda^* T^X_\lambda\;.
\end{equation}
Thus, the Theorem 
 allows us to use  (\ref{lemma2}) as a generating
formula for a gauge transformation to an arbitrary order $n$.
To second order, we have explicitly 
\begin{equation}
\label{eq:tgt} T^Y_\lambda=T^X_\lambda + 
\lambda\pounds_{\xi_{(1)}} T^X_\lambda
+\frac{\lambda^2}{2}\,\left(\pounds^2_{\xi_{(1)}}+
\pounds_{\xi_{(2)}}\right) T^X_\lambda
+\ldots\;,
\label{fund}
\end{equation}
where $\xi_{(1)}$ and $\xi_{(2)}$ are now the first two generators of
$\Phi_\lambda$, or of the gauge transformation, if one prefers.
We can now relate the perturbations in the two gauges.  To the
lowest orders, this is easy to do explicitly, just substituting (\ref{4.4})  
into   (\ref{eq:tgt}):

\noindent {\bf Proposition 2:} {\it Given a tensor field $T$,  
the relations between its first and second
order perturbations in two different gauges are:}
\begin{equation} 
\delta T^Y-\delta T^X=\pounds_{\xi_{(1)}}T_0\;;
\label{first}
\end{equation} 
\begin{equation}
\delta^2 T^Y-\delta^2 T^X=\left(\pounds_{\xi_{(2)}}+
\pounds^2_{\xi_{(1)}}\right)T_0 
+2\pounds_{\xi_{(1)}}\delta T^X \;.
\label{second}
\end{equation}

This result is consistent with Proposition 1, of course.  Equation
(\ref{first}) implies that $T_\lambda$ is gauge-invariant to the first
order iff $\pounds_\xi T_0=0$, for any vector field $\xi$ on $\cal M$.
In particular, one must have $\pounds_{\xi_{(2)}}T_0=0$, and therefore
Eq.\ (\ref{second}) leads to $\pounds_\xi\delta T=0$.  Similar
conditions hold at higher orders.
It is also possible to find the explicit expressions, in terms of $X$
and $Y$, for the generators $\xi_{(k)}$ of a gauge transformation. In
fact, it is easy to prove that the first two generators of the
one-parameter family of  diffeomorphisms $\Phi$ are 
$
\xi_{(1)}=Y-X$, and 
$\xi_{(2)}=[X,Y]$.


\section{Conclusions}
In this contribution we have briefly illustrated how a gauge
transformation in the theory of spacetime perturbations 
is a member of a family ---  not 
a group --- of diffeomorphisms. The family can be   approximated
by a flow only when attention is restricted to linear perturbations.
When $n$-th order perturbations are considered, gauge transformations
can instead be  approximated by $n$-th rank knights 
diffeomorphisms. In fact, in
introducing these objects, we have proved\cite{bi:BMMS} that a generic
family of diffeomorphisms can always be regarded  as a 
knight, in general   of $\infty$-th rank. 
We have also considered gauge invariance, 
and given a generating formula for gauge transformations
of arbitrary order. 
The formalism presented here has been applied
in the context of cosmology,\cite{bi:BMMS} 
to obtain the explicit
trasformations  between the syncronous and the
Poisson (generalized longitudinal) gauges.
\\ 

\noindent
{\bf Acknowledgements:}
This work has been partially supported by the Italian MURST; MB thanks
ICTP and INFN for financial support. 
 MB and SS are grateful to Dennis W.\ Sciama for hospitality at the
Astrophysics Sector of SISSA.


\begin{thebibliography}{99}

\bibitem{bi:MPS}
K.\ Tomita, Prog.\ Theor.\ Phys.\ {\bf 37}, 831 (1967);
S.\ Matarrese, O.\ Pantano, \& D.\ Saez, Phys.\
Rev.\ Lett.\ {\bf 72}, 320
(1994); MNRAS {\bf 271},  513 (1994).

\bibitem{bi:russ}
D.\ S.\ Salopek, J.\ M.\ Stewart, \& K.\ M.\
 Croudace, MNRAS {\bf 271}, 1005 (1994);
H.\ Russ, M.\ Morita, M.\ Kasai, \& G.\ B\"orner, 
 Phys.\ Rev.\ D {\bf 53}, 6881 (1996).

\bibitem{bi:gleiseretal}
R.\ J.\ Gleiser, C.\ O.\ Nicasio, R.\ H.\ Price, \&
J.\ Pullin, Class.\ Quantum Grav.\ {\bf 13}, L117 (1996); 
preprint gr-qc/9609022.  


\bibitem{bi:sachs} R.\ K.\ Sachs, in {\em Relativity, Groups, and
Topology\/}, edited by C.\ DeWitt and B.\ DeWitt (Gordon and Breach,
New York, 1964).

\bibitem{bi:SW}
J.\ M.\ Stewart \& M.\ Walker,  Proc.\ R.\ Soc.\ London {\bf A 341},
49
(1974).

\bibitem{bi:BMMS}
M.\ Bruni, S.\ Matarrese, S.\ Mollerach, \& S.\ Sonego,
preprint IC/96/174, SISSA--136/96/A, gr-qc/9609040, submitted for
publication.
                                                 

\bibitem{bi:bardeen}
J.\ M.\ Bardeen,   Phys.\ Rev.\ D {\bf 22}, 1882 (1980);
H.\ Kodama and M.\ Sasaki, Prog.\ Theor.\ Phys.\ Suppl.\ 
{\bf 78}, 1 (1984);
J.\ M.\ Stewart, Class.\ Quantum Grav.\ {\bf 7}, 1169 (1990);
V.\ F.\ Mukhanov, H.\ A.\ Feldman, and R.\ H.\ 
Brandenberger, Phys.\
Rep.\ {\bf 215}, 203 (1992);
R.\ Durrer, Fund.\ Cosmic Phys.\ {\bf 15}, 209 (1994).

\bibitem{bi:EB}
G.\ F.\ R.\ Ellis, \& M.\ Bruni, Phys.\  Rev.\
D {\bf  40}, 6, 1804--1818, (1989).                


\bibitem{bi:schouten}
J.\ A.\ Schouten, {\em Ricci-Calculus\/} (Springer, Berlin, 1954), p. 108.

\bibitem{bi:SB}
S.\ Sonego \& M.\ Bruni, 
(1996)
submitted for publication.




\end{thebibliography}
\end{document}

\end